\documentclass[12pt]{article}
\usepackage{amssymb, amsmath,booktabs}
\usepackage{graphicx}
\usepackage{hyperref}

\usepackage{authblk}
\usepackage{natbib}

\title{\bf  Joint Modeling of Two Stochastic Processes, with Application to Learning Hospitalization Dynamics from Wastewater Viral Concentrations}

\author[a,c]{K.\ Ken Peng}
\author[b]{Charmaine B.\ Dean\thanks{Corresponding author}}
\author[c]{Robert Delatolla}
\author[a]{X.\ Joan Hu}
\author[c]{Elizabeth Renouf}

\affil[a]{Department of Statistics and Actuarial Science, Simon Fraser University,\\
8888 University Dr, Burnaby, BC, V5A 1S6, Canada}

\affil[b]{Department of Statistics and Actuarial Science, University of Waterloo,\\
200 University Ave W, Waterloo, ON, N2L 3G1, Canada}

\affil[c]{Department of Civil Engineering, University of Ottawa,\\
Ottawa, ON, Canada}

\date{}
\begin{document}

\maketitle

\bigskip
\begin{abstract}
In the post-pandemic era of COVID-19, hospitalization remains a primary public health concern and wastewater surveillance has become an important tool for monitoring its dynamics at the level of community. However, there is usually no sufficient information to know the infection process that results in both wastewater viral signals and hospital admissions. That key challenge has motived a statistical framework proposed in this paper. We formulate the connection of overtime wastewater viral signals and hospitalization counts through a latent process of infection at the level of individual subject. We provide a strategy for accommodating aggregated data, a typical form of surveillance data. Moreover, we ease the conventional procedure of the statistical learning with the joint modeling using available information on the infection process, which can be under-reporting. A simulation study demonstrates that the proposed approach yields stable inference under different degrees of under-ascertainment. The COVID-19 surveillance data from Ottawa, Canada shows that the framework recovers coherent temporal patterns in infection prevalence and variant-specific hospitalization risk under several reporting assumptions.
\end{abstract}

\noindent%
{\it Keywords:} Environmental surveillance, indirect statistical learning, multi-state processes, pseudo-likelihood estimation, wastewater epidemiology
\vfill

\newpage
\section{Introduction}
\label{W3:intro}
The COVID-19 pandemic has underscored the need for reliable and scalable methods to monitor disease transmission at a population level. Traditional surveillance methods, such as case reporting and hospitalization records, are often delayed and subject to biases due to testing availability and healthcare seeking behavior \citep{starnini2021impact, sherratt2021exploring}. As an alternative, wastewater and environmental surveillance (WES) has gained significant attention as a cost-effective and a leading indicator for tracking infectious diseases in communities \citep{parkins2024wastewater}. By analyzing viral signals in wastewater, researchers can estimate trends in infection prevalence without relying on individual-level testing data \citep{holcomb2024estimating, pappu2025tracking, mohring2024estimating}. Throughout this paper, we refer to the wastewater viral signal as the measured concentration of SARS-CoV-2 RNA in wastewater samples (e.g., primary sludge), expressed in gene copies per milliliter. This quantity reflects an aggregated environmental signal arising from viral shedding by infected individuals in the contributing population.

The application of WES to COVID-19 has led to numerous insights into community transmission dynamics. Studies have shown that wastewater viral concentrations correlate with reported case counts and hospitalizations, often providing an early warning signal of surges in infections \citep{larsen2020tracking, peng2023exploration, hegazy2022understanding}. Several methodologies have been employed to model these relationships, including distributed lag models \citep{galani2022sars, peccia2020measurement, kaplan2021aligning, zulli2022predicting, schoen2022sars, xie2022rna, peng2023exploration, PENG2025100840}, time series models \citep{jeng2025forecasting, lai2025temporal, simone2024time}, and compartmental epidemiological models \citep{mohring2024estimating, meadows2025epidemiological, miyazawa2024wastewater}. While these methods have demonstrated the utility of WES, they often rely on strong assumptions about the direct relationship between wastewater viral signals and clinical metrics.

A central challenge in linking wastewater viral signals to hospitalizations is that their association is dynamic and variant-dependent, rather than stable or directly interpretable. The same wastewater viral signals may correspond to very different hospitalization burdens over time, reflecting changes in infection prevalence, individual-level viral shedding, variant-specific hospitalization risk, and the coexistence of multiple viral variants within the population at any given time, rather than a single dominant variant defined by a discrete transition or threshold \citep[e.g.,][]{d2022wastewater}. Modeling their association directly therefore conflates different processes and limits inferential interpretability. Addressing these challenges requires modeling infection as an intermediary process that governs both viral shedding and hospitalization risk at the individual level. However, surveillance data rarely support such modeling directly. In practice, individual-level information is unavailable and observations are aggregated at the community level. Moreover, even when infection information such as reported case counts is available, it typically underestimates true infections due to asymptomatic cases, limited testing capacity, and changing testing policies \citep{lau2021evaluating, albani2021covid}. These two limiting scenarios motivate a statistical framework capable of learning individual-level disease processes using only imperfect population-level surveillance data.

To simultaneously analyze multiple related outcomes or processes while accounting for their dependencies, joint modeling is widely used in biostatistics and epidemiology. Traditional joint modeling often links a longitudinal submodel for repeated measurements (e.g., biomarkers) with a survival or event time submodel for time-to-event outcomes through latent structures \citep{tsiatis2004joint}. A comprehensive overview of inference procedures for joint models is provided by \citet{wu2012analysis}. Common applications include linking biomarkers with survival outcomes, modeling disease progression under treatment, and combining heterogeneous data sources in clinical research, such as HIV-related studies \citep{wu2010joint, temesgen2018joint, luvanda2023joint} and cardiovascular disease studies \citep{zelelew2023joint, gilani2017anthropometric}. 

In an inferential setting in which individual-level processes are not directly observed and only aggregated summaries are available, a substantial literature has shown that aggregation can induce information loss and lead to overconfident inference if not properly accounted for \citep{lawless1984information, orcutt1968data, banks2020parameter}. Estimation under aggregation has been studied in a range of settings, including Markov processes \citep{davis2002estimating, kelton1987comparison}, gamma distributed processes \citep{chen2017estimation}, and spatio-temporal models \citep{nguyen2023estimation}. While some theoretical results suggest that aggregation may incur no asymptotic information loss under restrictive conditions, such assumptions are rarely met in practice. To the best of our knowledge, these issues have not been systematically examined in the context of wastewater-based infectious disease surveillance. 

In this work, we tailor a joint modeling framework to the wastewater surveillance setting, where the infection process is only partially observed and available data are aggregated at the population level. We formulate a parsimonious individual-level joint model in which infection mediates both wastewater viral shedding and hospitalization risk, and develop an estimation strategy that enables inference under two common data limitations in practice: the absence of individual-level observations and the systematic under-reporting of infection information. Rather than relying on a direct or deterministic relationship between wastewater signals and clinical outcomes, our approach treats both processes as conditionally dependent on a common latent infection process. Inference is achieved by combining information across these sources within a coherent statistical framework, rather than by assuming that any single data stream fully captures disease dynamics.

The remainder of this paper is structured as follows: \autoref{W3:method} presents our proposed modeling framework. \autoref{W3:learn} discuss the estimation procedure. \autoref{w3:simulation} reports a simulation study that evaluates the performance of the proposed estimation procedure and examines strategies for variance estimation. \autoref{W3:case} applies the model to Ottawa’s wastewater and hospitalization data, demonstrating its practical implementation and results. \autoref{W3:dis} concludes with a discussion of the implications of our findings, potential extensions, and future directions in wastewater-based disease surveillance.

\section{A Joint Modeling Framework}
\label{W3:method}

\subsection{Notation and Modeling}
\label{W3:method:model}
We consider a joint modeling framework for multiple stochastic processes arising in infectious disease surveillance. We define three processes evolving over continuous calendar time $t \in (0,E)$, where $E$ denotes the end of the study period.

The first process, $s(t)$, represents the infection status at time $t$.  This process is defined on a finite state space and may be multi-state, allowing different types of infection (e.g., infection variants). The second process, $h(t)$, denotes the hospitalization status at time $t$, which we treat as a binary-valued process indicating whether the individual is hospitalized. The third process, $w(t)$, represents a surveillance target at time $t$, such as a continuous measurement of viral load. In addition, we allow for time varying covariates $x(t)$ that may influence one or more of these processes.

The joint model is formulated by assuming that, conditional on the latent infection process and observed covariates, the surveillance target and hospitalization processes are conditionally independent. At each time $t$, the infection process $s(t)$ acts as the driving process that links the surveillance target $w(t)$ and the hospitalization process $h(t)$. Specifically, conditional on the infection status $s(t)$ and covariates $x(t)$, the distributions of $w(t)$ and $h(t)$ are specified through $[w(t)\mid s(t), x(t)]$ and $[h(t)\mid s(t), x(t)]$, respectively, while the evolution of $s(t)$ is governed by $[s(\cdot)\mid x(\cdot)]$. The joint model is defined through the conditional factorization
\begin{equation}
[s(\cdot), w(\cdot), h(\cdot) \mid x(\cdot)]
=
[s(\cdot) \mid x(\cdot)]
[w(\cdot) \mid s(\cdot), x(\cdot)]
[h(\cdot) \mid s(\cdot), x(\cdot)].
\label{jointmodel}
\end{equation}
This factorization is defined at the level of a generic individual and serves as the building block for population-level inference. Vaccination is an important determinant of both infection dynamics and hospitalization risk. While vaccination effects are not modeled explicitly here due to data limitations and aggregation, the proposed framework readily accommodates vaccination- or immunity-related covariates as a natural extension.

This formulation induces a joint model in which infection mediates the association between the two observable processes. Viral shedding, as captured by the surveillance target $w(t)$, and hospitalization risk, as captured by $h(t)$, are both driven by infection status, but represent distinct manifestations of disease progression. Conditional on $s(t)$, we assume that $w(t)$ and $h(t)$ are independent, reflecting the interpretation that infection status provides a sufficient summary of the aspects of disease severity relevant to both viral shedding and hospitalization at time $t$. This conditional independence assumption simplifies the joint structure while preserving the key mechanistic link through infection.

In practice, individual-level infection status and genetic variant are difficult to observe perfectly, and may be partially observed, delayed, or unobserved altogether. By contrast, surveillance targets such as viral load measurements and hospitalization events are more readily observed and may provide indirect information about infection status. For example, the presence of a positive viral signal or a hospitalization event is indicative of infection, even when infection status is not directly measured. The proposed framework is formulated at the individual level and is compatible with settings in which some components of the process are only partially observed. 

We now describe a general formulation for each component of the joint model in
\eqref{jointmodel}. The specific parametric choices introduced in
Section~\ref{W3:method:comp} are particular instances of this general framework.

\paragraph{Infection process.}
The latent infection process $s(\cdot)=\{s(t):t\in(0,E)\}$ is modeled as a
continuous-time multi-state stochastic process with finite state space
$\{0,1,\ldots,K\}$, where state $0$ denotes the uninfected state and states
$k=1,\ldots,K$ correspond to different genetic variants.
The evolution of $s(\cdot)$ is characterized through transition intensities
$\gamma_{jl}(t,\mathcal H_t)$, where $\mathcal H_t$ denotes the history of the
process up to time $t$.
This formulation accommodates both time-inhomogeneous Markov models, in which
$\gamma_{jl}(t,\mathcal H_t)$ depends only on the current state and calendar
time, as well as semi-Markov extensions that allow the transition intensities to
depend on the sojourn time in the current state.

\paragraph{Viral load process.}
The surveillance process $w(\cdot)$ represents an individual-level measurement
associated with infection, such as viral shedding.
Conditional on the latent infection state and covariates, $w(t)$ is specified
through a state-dependent conditional distribution,
\[
w(t)\mid s(t),\mathbf x(t)\sim G_{s(t)}\big(\cdot;\,\mathbf x(t)\big),
\]
where $\{G_k(\cdot;\mathbf x):k=0,\ldots,K\}$ denotes a family of distributions
indexed by infection status.
This formulation treats the surveillance signal as a measurement or emission
process driven by the underlying infection dynamics.
The choice of $G_k$ may vary depending on the application and data
characteristics, allowing for flexible distributional forms beyond the
illustrative gamma specification used in this work.

\paragraph{Hospitalization process.}
The hospitalization process is modeled as a time-to-event process describing
the time to first hospital admission.
Equivalently, it may be viewed as a two-state process with states
\{not hospitalized, hospitalized\}, where the hospitalized state is absorbing.
Let $H$ denote the hospitalization time, if it occurs.
Conditional on infection status and covariates, hospitalization risk is
characterized through a hazard function,
\[
\lambda(t\mid s(t-),\mathbf x(t)),
\]
defined on the calendar time scale.
Mechanistically, hospitalization risk depends on the elapsed time since
infection; this dependence is implicitly captured through the infection
process and the infection onset times $\{T_{k,r}\}$.
The calendar-time formulation adopted here is therefore equivalent to a
duration-since-infection representation, while remaining compatible with
population-level aggregated data.

\subsection{Specifications of Model Components}
\label{W3:method:comp}

The general framework described in Section~\ref{W3:method:model} admits a wide
range of modeling choices for the infection, surveillance, and hospitalization
processes.
In this section, we present the specific parametric specifications adopted in
our analysis.
These choices are motivated by the structure of the available wastewater and
hospitalization data and are intended to provide a parsimonious yet flexible
representation of epidemic dynamics.

\subsubsection{Infection process $s(\cdot)\mid x(\cdot)$}
\label{W3:method:comp:s}

As a example of the general multi-state infection model introduced in
Section~\ref{W3:method:model}, we consider a time-inhomogeneous Markov formulation
that allows for multiple genetic variants.
Transitions are restricted to occur between the uninfected state and infected
states, thereby excluding co-infection and direct transitions between different
genetic variants. At any given time $t$, multiple variant-specific infection intensities $\gamma_{0k}(t)$ may be positive, allowing for the co-circulation of different genetic variants at the population level.

The transition intensity
from the uninfected state to genetic variant $k$ is modeled using a mixture of
Gaussian functions,
\begin{equation}
\gamma_{0k}(t)
=
\sum_{m \in \mathcal{M}_k}
a_{k,m}
\exp\!\left\{-\frac{(t-b_{k,m})^2}{2c_{k,m}^2}\right\},
\qquad k=1,\ldots,K.
\end{equation}
Here, $\mathcal{M}_k$ indexes the set of Gaussian components used to represent the temporal incidence profile of variant $k$, with the number of components chosen to reflect the number of distinct epidemic periods considered for that variant. This specification flexibly represents the rise and decline of infection incidence over time. While covariate effects are not included in the infection process for simplicity, the general framework readily accommodates such extensions if relevant information is available. Gaussian components provide a smooth and parsimonious representation of epidemic waves, while mixtures allow multiple waves or periods of transmission to be captured without imposing abrupt transitions. This choice offers a flexible yet interpretable approximation to time-varying infection incidence.

\subsubsection{Viral load process $w(\cdot)\mid s(\cdot),\mathbf{x}(\cdot)$}
\label{W3:method:comp:w}

We specify the viral load process as a state-dependent measurement model
conditional on infection status and covariates.
In particular, we instantiate the general family $G_k(\cdot;\mathbf{x})$
introduced above using a gamma distribution when infection is present.
Motivated by the nonnegativity and right-skewness of viral load
measurements, we assume a gamma distribution for the surveillance signal when infection is present,
\begin{equation}
w(t)\mid s(t),\mathbf{x}(t)
\sim
\begin{cases}
0, & s(t)=0, \\
\mathrm{Gamma}\!\left(\alpha_k,\,
\exp\{\beta^{(k)}_0+
\mathbf{x}(t)^\top\boldsymbol{\beta}^{(k)}\}\right), & s(t)=k,
\end{cases}
\end{equation}
This specification provides a flexible mean structure while allowing for
substantial variability in viral shedding across individuals.
Alternative distributional choices could be accommodated within the same
framework if warranted by the data.

\subsubsection{Hospitalization process $h(\cdot)\mid s(\cdot),\mathbf{x}(\cdot)$}
\label{W3:method:comp:h}

Although hospitalization is conceptually represented as a binary state (hospitalized versus not hospitalized), the available hospitalization data record only admissions and not discharge information. Accordingly, we consider a simplified case and model hospitalization as the time to first hospital admission. This corresponds to a two-state process in which individuals transition from a non-hospitalized state to an absorbing hospitalized state.

Conditional on infection status, we specify the hazard of hospitalization on the
calendar time scale as
\begin{equation}
\lambda(t\mid s(t-),\mathbf{x}(t))
=
\begin{cases}
0, & s(t-)=0, \\
\exp\{\lambda^{(k)}_0+
\mathbf{x}(t)^\top\boldsymbol{\lambda}^{(k)}\}, & s(t-)=k,
\end{cases}
\end{equation}
Under this specification, $\exp\{\lambda^{(k)}_0\}$ represents a variant-specific baseline hazard, which is assumed to be constant over time for individuals infected with variant $k$, conditional on the covariate $\mathbf{x}(t)$. The hazard is formulated on the calendar time scale to maintain consistency with the available aggregated surveillance data, which do not provide information on individual-level times since infection. Allowing baseline hazard to vary with the elapsed time since infection would require introducing an alternative time scale based on infection onset, which is not pursued here.

\section{Statistical Learning}
\label{W3:learn}
\subsection{Inference under Complete Individual-Level Data}
\label{W3:learn:indi}

We first describe likelihood-based inference under an idealized setting in which complete individual-level information is available. This formulation serves as a conceptual reference point and provides a clear link between the individual-level joint model introduced in Section~\ref{W3:method} and the aggregated-data estimation strategy developed in subsequent sections.

We consider a population of $N$ individuals indexed by $i=1,\ldots,N$. For individual $i$, let $s_i(t)$, $w_i(t)$, and $h_i(t)$ denote the infection status, surveillance measurement, and hospitalization process at time $t\in(0,E)$, respectively, with associated covariate process $\mathbf{x}_i(t)$. Throughout this subsection, we assume that the full trajectories $\{s_i(t), w_i(t), h_i(t): t\in(0,E)\}$ are observed without error. All individuals are assumed to start in the uninfected state, $s_i(0)=0$.

For individual $i$, let $T_i$ denote the time to first hospitalization and let $E$ denote the end of the study period. The observed follow up time is $U_i = T_i \wedge E$, with censoring indicator $\delta_i = I(T_i < E)$. In addition, suppose that the individual-level surveillance process $\{w_i(t)\}$ is observed without missingness at a discrete set of observation times (e.g., daily), and that the infection process $\{s_i(t)\}$ is fully observed at the individual level.

Let parameter vectors
$\boldsymbol{\theta}_h$, $\boldsymbol{\theta}_w$, and $\boldsymbol{\theta}_s$,
which govern the hospitalization, surveillance, and infection processes related parameters, respectively. Conditional on the infection process, the likelihood contribution of the hospitalization process is given by
\begin{equation}
L_{h,i}(\pmb{\theta}_h)
=
\lambda\!\left(U_i \mid s_i(U_i-), \mathbf{x}_i(U_i)\right)^{\delta_i}
\exp\!\left\{-\int_0^{U_i}
\lambda\!\left(v \mid s_i(v-), \mathbf{x}_i(v)\right)\, dv \right\},
\label{eq:h_likeli}
\end{equation}
where $\lambda(t \mid s_i(t-), \mathbf{x}_i(t))$ denotes the conditional hazard of hospitalization.

Let $\mathcal{T}_i\subset(0,E)$ denote the set of observation times at which the surveillance target $w_i(t)$ is measured (e.g., daily measurements). Under the conditional model $[w_i(t)\mid s_i(t),\mathbf{x}_i(t)]$ with density $f_w(\cdot\mid s_i(t),\mathbf{x}_i(t))$, the surveillance likelihood is
\begin{equation}
\label{w_likeli}
L_{w,i}(\pmb{\theta}_w)
=
\prod_{t\in\mathcal{T}_i}
f_w\!\big(w_i(t)\mid s_i(t),\mathbf{x}_i(t)\big).
\end{equation}
For the illustrative gamma specification in \eqref{W3:method:comp:w}, this density corresponds to a point mass at $0$ when $s_i(t)=0$ and a gamma density when $s_i(t)=k$.

The infection process is modeled as a continuous time multi-state process with transition intensities $\gamma_{jl}(t)$. Let $0 = t_{i0} < t_{i1} < \cdots < t_{iJ_i} < E$ denote the transition times of $s_i(\cdot)$, and let $s_{ij}$ denote the state occupied on $(t_{ij}, t_{i,j+1})$. The likelihood contribution of the infection process is
\begin{equation}
\label{s_likeli}
L_{s,i}(\pmb{\theta}_s)
=
\left\{
\prod_{j=1}^{J_i}
\gamma_{s_{i,j-1},\,s_{ij}}\!\left(t_{ij},\mathcal{H}_{i t_{ij}}\right)
\right\}
\exp\!\left\{
-\int_{0}^{E}
\Gamma_{s_i(v)}\!\left(v,\mathcal{H}_{iv}\right)\,dv
\right\},
\end{equation}
where $\Gamma_{s_i(v)}(v,\mathcal{H}_{iv})=\sum_{l\neq r}\gamma_{s_i(v) l}(v,\mathcal{H}_{iv})$ is the total exit intensity from state $s_i(v)$ at time $v$. Under the simplified transition structure used in our application (transitions only between state $0$ and states $1,\ldots,K$), the sum in $\Gamma_{s_i(v)}(\cdot)$ is correspondingly restricted. 

If $s_i(\cdot)$, $w_i(\cdot)$, and $h_i(\cdot)$ were all observed at the individual level, then under the conditional independence assumption $h_i(\cdot)\perp w_i(\cdot)\mid s_i(\cdot),x_i(\cdot)$, the complete data likelihood factorizes as
\begin{equation*}
L_{i}^{(c)}(\pmb{\theta}_s,\pmb{\theta}_w,\pmb{\theta}_h)
=
L_{h,i}(\pmb{\theta}_h)\,
L_{w,i}(\pmb{\theta}_w)\,
L_{s,i}(\pmb{\theta}_s).
\end{equation*}
If the infection process $s_i(\cdot)$ is unobserved, inference is based on the marginal likelihood obtained by integrating out the latent path,
\begin{equation*}
\label{marginal_likeli}
L_{i}^{(o)}(\pmb{\theta}_s,\pmb{\theta}_w,\pmb{\theta}_h)
=
\int_{\mathcal{S}}
L_{h,i}(\pmb{\theta}_h)\,
L_{w,i}(\pmb{\theta}_w)\,
L_{s,i}(\pmb{\theta}_s)\,
d[s_i(\cdot)],
\end{equation*}
where $\mathcal{S}$ denotes the space of admissible infection trajectories. Assuming independence across individuals, the full marginal likelihood is
\begin{equation*}
\label{overall_marginal_likeli}
L^{(o)}(\pmb{\theta}_s,\pmb{\theta}_w,\pmb{\theta}_h)
=
\prod_{i=1}^{N}
L_{i}^{(o)}(\pmb{\theta}_s,\pmb{\theta}_w,\pmb{\theta}_h).
\end{equation*}
In practice, surveillance systems typically provide only aggregated summaries of $w_i(\cdot)$ and $h_i(\cdot)$ rather than individual-level trajectories, motivating the pseudo likelihood approach developed next.

\subsection{Pseudo Likelihood under Aggregated Surveillance Data}
\label{W3:learn:est}
In practice, complete individual-level trajectories of infection status and viral shedding are rarely observed in surveillance systems. Instead, available data typically consist of population-level summaries collected at discrete time points. This motivates an estimation strategy that links the individual-level joint model introduced in ~\autoref{W3:method} to aggregated observations through a pseudo likelihood construction.

We consider observations indexed by discrete time $t=0,1,\ldots,E$. Let $H_t$ denote the total number of new hospital admissions associated with target disease at time $t$, and let $W_t^{(k)}$ denote the aggregated surveillance signal associated with genetic variant $k$, obtained by summing individual-level contributions,
\[
H_t = \sum_{i=1}^N h_i(t),
\qquad
W_t^{(k)} = \sum_{i=1}^N w_i(t)\, I\{s_i(t)=k\}.
\]
In addition, surveillance systems may reported an observed number of active cases $S_t^*$, which generally underestimates the true number of infected individuals
\[
S_t = \sum_{i=1}^N I\{s_i(t)\neq 0\}.
\]
We do not assume that $S_t^*$ is an unbiased or complete measure of $S_t$, but treat it as partial information about infection prevalence. Different assumptions about the degree of case ascertainment lead to different interpretations of $S_t^*$ within the pseudo-likelihood, and these assumptions are explored systematically through a scenario analysis in ~\autoref{W3:case:scena}.

To relate the individual-level model to aggregated data, we discretize time and approximate the continuous time likelihood by its discrete time counterpart. Under independence across individuals, the individual hospitalization likelihood implies that, conditional on genetic variant, daily hospital admissions follow a Poisson type likelihood. Specifically, letting $H_t^{(k)}$ and $R_t^{(k)}$ denote the (unobserved) number of hospitalizations and individuals at risk associated with genetic variant $k$ at time $t$, the hospitalization contribution takes the form
\begin{equation}
L_h(\boldsymbol{\theta}_h)
=
\prod_{t=0}^E
\prod_{k=1}^K
\Big[
f(\mathbf{X}_t;\boldsymbol{\lambda}^{(k)})^{H_t^{(k)}}
\exp\!\big\{-f(\mathbf{X}_t;\boldsymbol{\lambda}^{(k)})\,R_t^{(k)}\big\}
\Big],
\label{eq:Ht_pseudo}
\end{equation}
where $\mathbf{X}_t$ denotes population level covariates.

Similarly, under the conditional gamma model for individual surveillance measurements, the aggregated signal $W_t^{(k)}$ follows a gamma distribution with shape proportional to the number of infected individuals of type $k$,
\begin{equation}
L_w(\boldsymbol{\theta}_w)
=
\prod_{t=0}^E
\prod_{k=1}^K
g\!\left(
W_t^{(k)};
S_t^{(k)}\alpha_k,\,
f(\mathbf{X}_t;\boldsymbol{\beta}^{(k)})
\right),
\label{eq:Wt_pseudo}
\end{equation}
where $g(\cdot)$ denotes the gamma density and $S_t^{(k)}$ is the number of individuals infected with type $k$ at time $t$.

\paragraph{Expectation based approximation.}
The quantities $H_t^{(k)}$, $R_t^{(k)}$, and $S_t^{(k)}$ are not directly observed. To construct a tractable likelihood, we replace these latent variant-specific quantities by their conditional expectations given the aggregated observations. Specifically, we decompose population-level counts according to the infection composition at time $t$,
\begin{align*}
E\!\left(S_t^{(k)}\right) &= S_t \, P\!\left\{s(t)=k \mid s(t)\neq 0\right\},\\
E\!\left(H_t^{(k)}\right) &= H_t \, P\!\left\{s(t)=k \mid s(t)\neq 0\right\},\\
E\!\left(R_t^{(k)}\right) &= E\!\left(S_t^{(k)}\right)\left(1-\frac{C_t}{N}\right),
\qquad C_t=\sum_{u=0}^t H_u,
\end{align*}
where
\[
P\!\left\{s(t)=k \mid s(t)\neq 0\right\}
=
\frac{P\!\left(s(t)=k\right)}{\sum_{l=1}^K P\!\left\{s(t)=l\right\}}.
\]
The approximation for $E(R_t^{(k)})$ implicitly assumes that prior hospitalization history is approximately independent of current infection status and variant type at the population level. Under this mean-field assumption, $1-C_t/N$ serves as a proxy for the fraction of individuals remaining at risk for a first admission at time $t$, which is then applied proportionally to the expected number of individuals currently infected with variant $k$.

The probabilities $P\{s(t)=k\}$ are determined by the infection process model through its transition intensity structure. In particular, letting
\[
\boldsymbol{\rho}(t;\boldsymbol{\theta}_s)
=
\big(\rho_1(t;\boldsymbol{\theta}_s),\ldots,\rho_K(t;\boldsymbol{\theta}_s)\big)
\]
denote the vector of state occupancy probabilities at time $t$, we have
\[
P\{s(t)=k\} = \rho_k(t;\boldsymbol{\theta}_s),
\]
where $\boldsymbol{\rho}(t;\boldsymbol{\theta}_s)$ is obtained by propagating the initial state distribution forward in time using a discrete time approximation of the continuous time transition intensity matrix.

When only a lower bound $S_t^*$ on $S_t$ is available, we approximate $S_t$ by its conditional expectation under a binomial model,
\[
S_t \sim \text{Binomial}\{N, P(s(t)\neq 0)\},
\]
leading to
\[
E(S_t \mid S_t \geq S_t^*) \approx
\mu_s(t) +
\sigma_s(t)
\frac{\phi((S_t^*-\mu_s(t))/\sigma_s(t))}{1-\Phi((S_t^*-\mu_s(t))/\sigma_s(t))},
\]
where $\mu_s(t) = N P(s(t)\neq 0)$ and $\sigma^2_s(t) = N P(s(t)\neq 0)\{1-P(s(t)\neq 0)\}$. Here $\phi(\cdot)$ and $\Phi(\cdot)$ denote the probability density function and cumulative distribution function of the standard normal distribution, respectively, arising from a normal approximation to the binomial distribution.

\paragraph{Pseudo likelihood.} Substituting these expectation based approximations yields the \textit{pseudo likelihood}:
\begin{align*}
L^{\mathrm{PL}}(\boldsymbol{\theta}_s,\boldsymbol{\theta}_w,\boldsymbol{\theta}_h)
=
\prod_{t=0}^E
L_h\!\left(E(H_t^{(k)}),E(R_t^{(k)});\boldsymbol{\theta}_h\right)
L_w\!\left(E(S_t^{(k)});\boldsymbol{\theta}_w\right)
L_{S^*}\!\left(S_t^*;\rho(t;\boldsymbol{\theta}_s)\right),
\end{align*}
where all latent variant-specific quantities are replaced by their conditional expectations and the dependence on $\boldsymbol{\theta}_s$ enters through the infection state probabilities $\rho_k(t;\boldsymbol{\theta}_s)$. The component $L_{S^*}(S_t^*;\rho(t;\boldsymbol{\theta}_s))$ is constructed from a
binomial working model for the latent number of infected individuals
$S_t \sim \mathrm{Binomial}\!\left(N, \sum_{k=1}^K \rho_k(t;\boldsymbol{\theta}_s)\right)$.
Because reported case counts $S_t^*$ systematically underestimate true infections,
we treat $S_t^*$ as providing a one sided constraint rather than an exact observation.
Specifically, when reported case counts are treated as underreported, the contribution takes the form
\[
L_{S^*}(S_t^*;\rho(t;\boldsymbol{\theta}_s))
=
\Pr(S_t \geq S_t^* \mid \rho(t;\boldsymbol{\theta}_s)),
\]
while in periods where reported cases are assumed complete, the usual binomial
probability mass function is used. This construction allows reported case counts
to inform infection prevalence without being interpreted as unbiased realizations of the true infection process.

Parameter estimates are obtained by maximizing the resulting pseudo log likelihood. While the observed Fisher information of the pseudo likelihood can be used for variance estimation, our simulation study in ~\autoref{w3:simulation} shows that it tends to underestimate uncertainty. We therefore recommend a parametric bootstrap procedure for uncertainty quantification, as empirical results in Section~\autoref{w3:simulation} indicate that it provides stable uncertainty estimates under the aggregation and approximation schemes considered here. In each bootstrap replicate, individual-level data are generated from the fitted model, re-aggregated to the observed data scale, and refitted; the resulting empirical variability of parameter estimates is used to approximate standard errors. The data-generating mechanism mirrors that used in the simulation study, with model parameters fixed at their fitted values (see Supplementary Material).

\section{Simulation Study}
\label{w3:simulation}

\subsection{Motivation and Design}
\label{w3:simulation:moti}

The simulation study evaluates the finite sample performance of the proposed pseudo likelihood approach under realistic surveillance constraints. Two challenges motivate the design. First, individual-level infection trajectories are unobserved and inference must rely on aggregated summaries. Second, reported case counts may substantially underestimate true infections due to incomplete testing and changing surveillance practices.

To reflect these features, data are generated from the individual-level joint model described in ~\autoref{W3:method}, while estimation is performed using only aggregated quantities. Under-reporting is introduced through two parameters. The parameter $r_1$ denotes the proportion of time points at which reported case counts can be treated as approximately complete, while $1-r_1$ corresponds to periods subject to under-reporting. The parameter $r_2$ denotes the reporting rate among infected individuals during under-reported periods. This setup allows us to examine a range of surveillance scenarios, from settings with frequent and reliable case reporting to situations characterized by prolonged reporting limitations and severe under-ascertainment.

\subsection{Simulation Setup}
\label{w3:simulation:setup}

We consider a closed population of size $N=100{,}000$ observed over a study period of length $E=200$ days, with $K=2$ genetic variants. Individual-level data are generated from the joint model described in ~\autoref{W3:method}, consisting of a multi-state infection process, a conditional viral shedding process, and an infection-dependent hospitalization hazard. Parameter values are chosen to induce two partially overlapping epidemic waves with distinct hospitalization risks.

To assess the impact of under-reporting, we consider four surveillance scenarios indexed by $(r_1,r_2)$. The parameter $r_1 \in \{0.2, 0.8\}$ denotes the proportion of time points at which reported case counts are approximately complete, while $r_2 \in \{0.2, 0.8\}$ denotes the reporting rate among infected individuals during under-reported periods. This yields four combinations,
\[
(r_1,r_2) \in \{(0.2,0.2), (0.2,0.8), (0.8,0.2), (0.8,0.8)\},
\]
ranging from mild to severe under ascertainment of infection prevalence. These values are chosen to span a range of plausible surveillance conditions, from settings with largely accessible testing and high reporting to scenarios characterized by prolonged testing limitations and low individual-level reporting. 

For each scenario, we generate 200 independent simulation replications. Within each replication, individual-level data are simulated and then aggregated to obtain daily case counts, hospital admissions, and wastewater viral signals. Model parameters are subsequently estimated using only these aggregated summaries. Full details of the data generating mechanisms and parameter values are provided in Supplementary Material~S1.

\subsection{Estimation Procedure}
\label{w3:simulation:proce}

For each simulated dataset, parameters are estimated under two approaches:
(i) a naive approach applies the pseudo likelihood framework as in ~\autoref{W3:learn:est}, but treats the reported case counts $S_t^*$ as if they were equal to the true number of infected individuals $S_t$ at all time points. That is, no allowance is made for under-reporting, and $S_t^*$ is directly substituted for $S_t$ in the likelihood construction and
(ii) the proposed pseudo likelihood approach that explicitly accounts for under-reporting.

For each approach, uncertainty is quantified using two methods: standard errors based on the inverse observed Fisher information of the pseudo likelihood, and parametric bootstrap standard errors. The bootstrap procedure is based on 200 resampled datasets generated from the fitted model.

\subsection{Results}
\label{w3:simulation:result}

Results are summarized over 200 simulation replications. ~\autoref{tabw3:simulation} reports, for each under-reporting scenario, the average parameter estimates and empirical standard deviations. Results are shown for both the naive approach and the proposed method, along with two standard error estimates.

Across all scenarios, ignoring under-reporting leads to systematic bias in infection and hospitalization parameters. The bias is modest under mild under-reporting but becomes substantial as either the frequency ($r_1$) or severity ($r_2$) of under-reporting increases. In contrast, the proposed pseudo likelihood approach yields accurate parameter recovery across all scenarios, including the most severe under-reporting setting. Notably, performance under the scenario $(r_1,r_2)=(0.8,0.8)$, which corresponds to mild under-reporting and near-complete case ascertainment, is comparable to that observed under more severe under-reporting regimes. This suggests that the proposed approach remains stable even when under-reporting is minimal, and does not degrade when the true reporting process is closer to complete observation.

Regarding uncertainty quantification, standard errors based on the inverse Fisher information matrix consistently underestimate the empirical variability of the estimators. Parametric bootstrap standard errors closely match the empirical standard deviations across all parameters and scenarios. Based on these findings, bootstrap-based inference is adopted in the subsequent real data analysis.

\section{Case Study: COVID-19 Surveillance in Ottawa}
\label{W3:case}

We apply the proposed joint modeling framework to COVID-19 surveillance data from Ottawa, Canada, with the goal of illustrating the dynamic association between wastewater viral signals and hospitalization records linked through latent infection dynamics under incomplete case ascertainment. The analysis focuses on the period spanning the transition from pre-Omicron variants to Omicron and its sub-variants, during which clinical testing practices and reporting completeness changed substantially.

\begin{figure}[!htp]
     \centering		\includegraphics[width=0.9\linewidth]{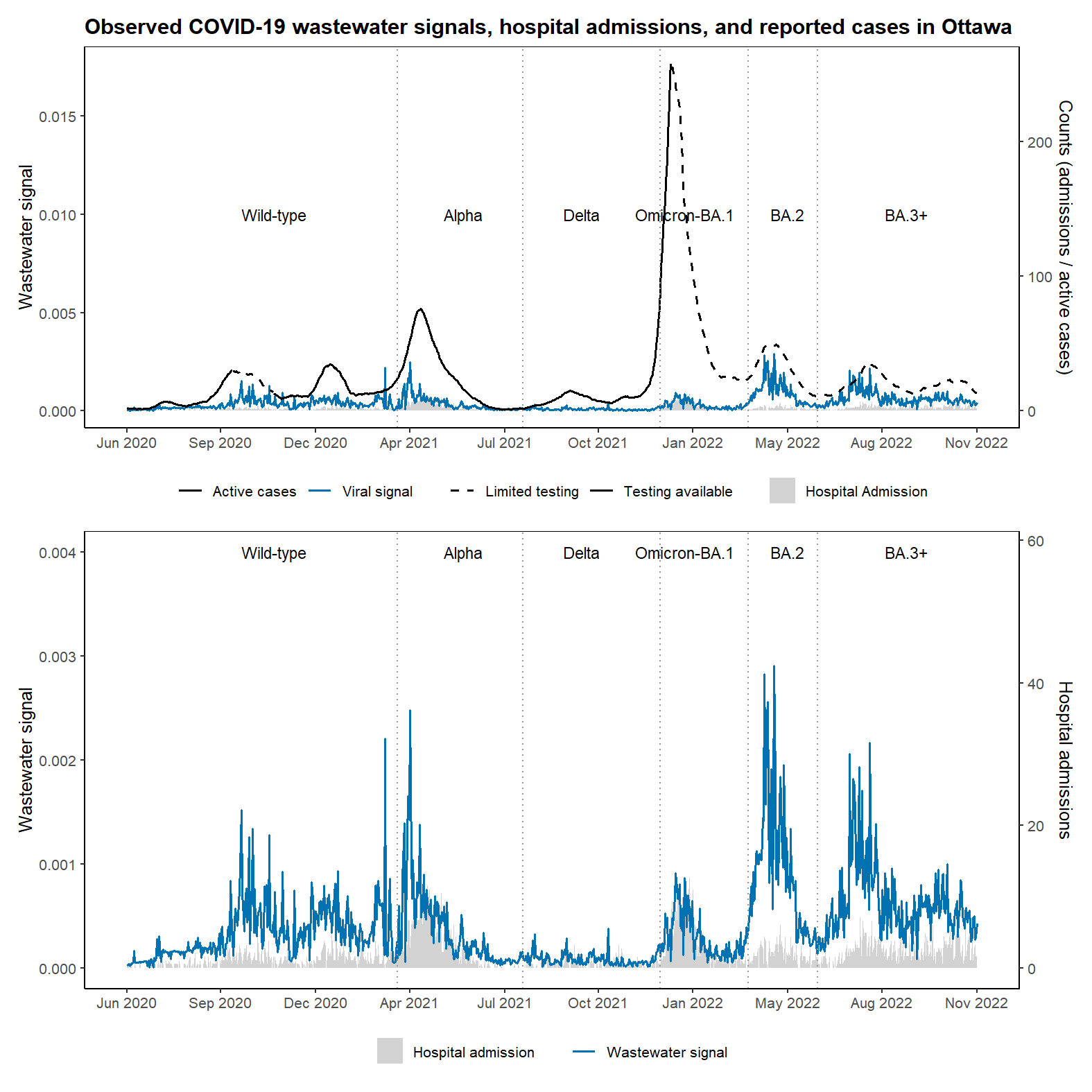}
		\caption{Wastewater viral signals, daily new COVID-19 hospital admissions, and reported active COVID-19 cases in Ottawa over the study period. Active case counts are rescaled to 1\% of their original values to facilitate comparison with wastewater signals and hospital admissions. Dashed segments of the active case curve indicate periods when testing access was limited and reported cases are treated as underreported in the scenario analysis.}
  \label{W3rawdata}
		\end{figure}

The available surveillance data consist of three aggregated time series observed at the population level: daily wastewater viral signals, daily new COVID-19, related hospital admissions, and reported active case counts. ~\autoref{W3rawdata} displays these three series over time, corresponding to the observed realizations of the surveillance, hospitalization, and reported infection processes, respectively. The dashed segments of the reported case curve indicate periods of limited testing accessibility, while vertical dashed lines mark the population-level period indicators used in the analysis, which are defined based on variant dominance inferred from wastewater allelic proportions following established criteria in the literature \citep{graber2021near}. The detailed definition of these periods, including specific dates and thresholds, is provided in the Supplementary Material.

Throughout the case study, we consider two major variant groups in the baseline analysis, indexed by $k=1,2$, corresponding to pre-Omicron variants and Omicron (including early sub-variants), respectively. Specifically, the pre-Omicron group includes the wild-type, Alpha, and Delta variants, while the Omicron group comprises Omicron BA.1, BA.2, and BA.3+. The temporal evolution of variant composition is constrained using known dominance periods inferred from surveillance, while allowing for transitional windows during which multiple variants may co circulate. Details of the variant mixture specification are provided in ~\autoref{W3:method} and Supplementary Material~S2.

Because individual-level recovery times are not observed in the available surveillance data, the transition intensity from infection to recovery, $\gamma_{k0}$, cannot be separately identified from the infection intensity. We therefore fix $\gamma_{k0}=0.07$ for all variant groups to impose a necessary identifiability constraint. This value corresponds to an average infectious duration of approximately 14 days, which is consistent with commonly reported estimates for SARS-CoV-2 infectious periods \citep{he2020temporal}. Importantly, fixing $\gamma_{k0}$ primarily determines the overall magnitude of the latent infection process, while the temporal patterns and the estimated relationships between the infections, wastewater viral signals, and hospital admissions remain largely unchanged.

While wastewater signals and hospital admissions are routinely collected throughout the study period, reported case counts are known to substantially underestimate true infections, particularly following changes in testing policy. To reflect this reality, our analysis explicitly accounts for uncertainty in case ascertainment rather than treating reported cases as unbiased measurements of infection prevalence.

\subsection{Scenario Analysis for Case Ascertainment}
\label{W3:case:scena}

One central challenge in linking wastewater viral signals and hospitalization data to underlying infection dynamics is uncertainty in reported active case counts. Changes in testing accessibility and eligibility over time imply that reported cases often represent a lower bound on true infections rather than a complete enumeration. To assess the impact of this uncertainty on inference, we conduct a scenario analysis that formalizes different assumptions about the relationship between reported active cases $S_t^*$ and the true number of infected individuals $S_t$.

We introduce a binary indicator $L(t)$ that encodes periods of broad testing accessibility. When $L(t)=1$, testing is assumed to be widely available and reported cases are treated as approximately complete; when $L(t)=0$ (corresponding to the dashed segments of the reported case curve in ~\autoref{W3rawdata}), reported cases are assumed to undercount true infections. This indicator is constructed from publicly available information on testing policy changes in Ontario and is used solely to define the ascertainment scenarios in ~\autoref{W3:case:scena}, rather than as a direct modeling input.

Three scenarios are considered:
\begin{itemize}
    \item \textbf{Scenario 1 (Full observability).} Reported active cases are assumed to equal true infections at all times, $S_t^* = S_t$. This scenario serves as a benchmark corresponding to analyses that ignore under-reporting.
    
    \item \textbf{Scenario 2 (Policy informed observability).} Reported cases are assumed to be complete only during periods of broad testing access:
    \[
    S_t^* = S_t \ \text{if } L(t)=1, \qquad S_t^* < S_t \ \text{if } L(t)=0.
    \]
    This scenario incorporates external information on testing policy.
    
    \item \textbf{Scenario 3 (Persistent under-reporting).} Reported cases are assumed to under estimate true infections throughout the study period, $S_t^* < S_t$ for all $t$. This scenario represents the most conservative setting.
\end{itemize}

Under each scenario, the same pseudo likelihood estimation procedure in ~\autoref{W3:learn} is used; scenarios differ only in how reported case counts are interpreted as information about the latent infection prevalence. Scenario~1 corresponds to the special case in which reported cases are treated as fully observed infections ($S_t^*=S_t$), whereas Scenarios~2 and~3 relax this assumption by treating $S_t^*$ as a lower bound on $S_t$ over policy-defined periods or throughout the study period, respectively. As demonstrated in the simulation study (~\autoref{w3:simulation}), the proposed approach remains stable even when reliable case information is available only intermittently.

\paragraph{Additional variant groupings.}
To demonstrate model flexibility, we also fit alternative variant mixture structures that further distinguish Omicron subvariants (e.g., separating BA.1 from later Omicron lineages). The corresponding specifications and results are reported in Supplementary Material~S2.

\subsection{Real data analysis results}
\subsubsection{Estimated infection prevalence and variant composition}
\label{W3:case:results:infection}

\begin{figure}
     \centering
		\includegraphics[width=0.8\linewidth]{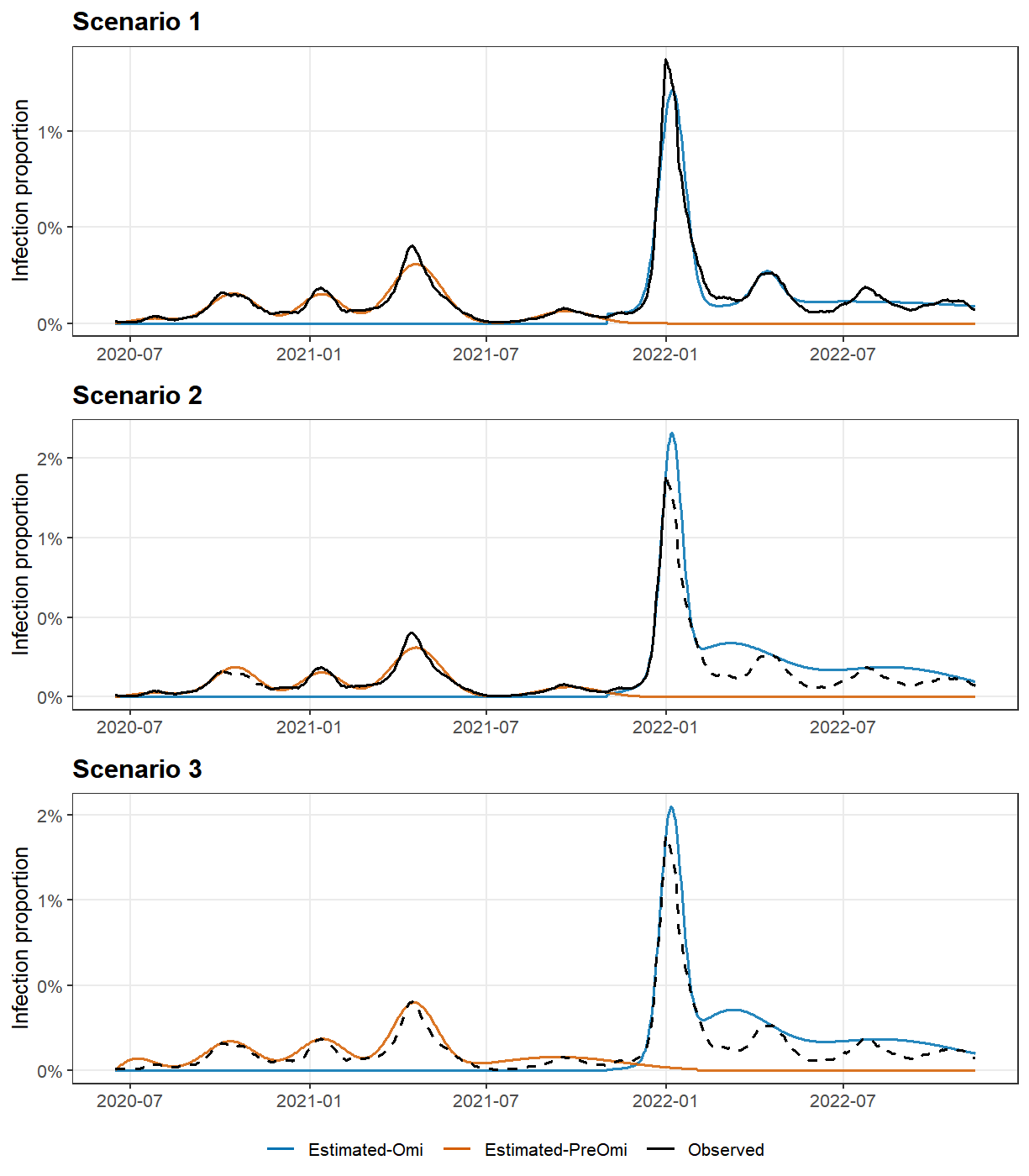}
		\caption{Estimated infection prevalence over time under the three case ascertainment scenarios described in Section~\ref{W3:case:scena}, together with reported active case counts. Lines are color-coded for the two variant groups, showing periods of co-circulation. Dashed segments of the reported case curve indicate periods treated as under-reporting.}
  \label{scenario_prop}
		\end{figure}

~\autoref{scenario_prop} displays the estimated infection prevalence over time under the three case ascertainment scenarios, together with reported active case counts. Results are stratified by the two variant groups, allowing for periods of co-circulation. Dashed segments of the reported case curve correspond to periods treated as under-reporting in Scenarios~2 and~3, as defined in ~\autoref{W3:case:scena}.

Across all scenarios, the inferred infection proportions capture the timing and overall shape of the observed epidemic waves. During transitional periods, the estimated state probabilities indicate concurrent circulation of multiple variants, with smooth transitions rather than abrupt replacement.

Differences across ascertainment scenarios primarily affect the estimated level of infection prevalence, while temporal patterns remain stable. When reported cases are treated as underreported (Scenarios~2 and~3), inferred infection prevalence is consistently higher than the observed counts, while preserving the timing and relative magnitude of epidemic peaks. Assuming complete case ascertainment (Scenario~1) yields estimates that more closely align with reported data.

\begin{figure}[!htp]
     \centering
		\includegraphics[width=0.8\linewidth]{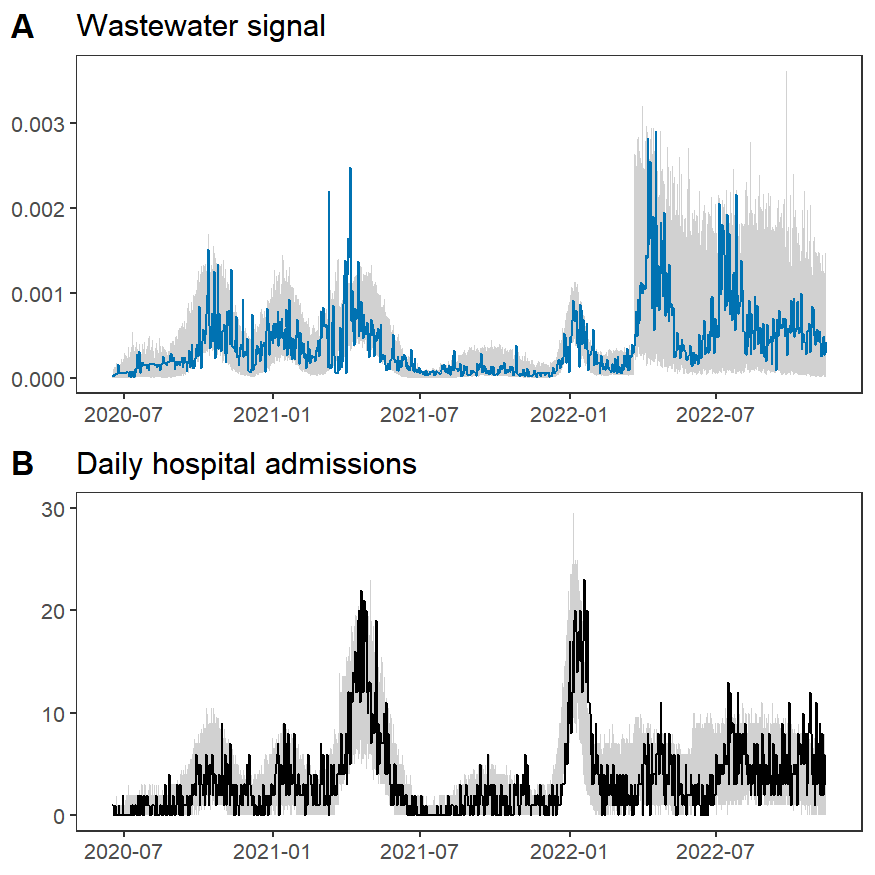}
		\caption{Observed wastewater viral signals and daily hospital admissions compared with trajectories simulated from the fitted joint model under the policy-informed case ascertainment scenario (Scenario~2). Shaded bands represent the 95\% central range of 100 simulated trajectories generated conditional on the fitted parameters, reflecting variability induced by the observation models rather than inferential uncertainty.}
  \label{fit_check}
		\end{figure}

~\autoref{fit_check} provides a complementary assessment by comparing observed wastewater viral signals and daily hospital admissions with trajectories simulated from the fitted model under the policy-informed scenario (Scenario~2). For each outcome, 100 replicate datasets are generated using the fitted parameter estimates. In both cases, the observed time series fall within the 95\% central range of the simulated trajectories, indicating that the joint model adequately captures the magnitude, temporal variability, and major peaks of both surveillance processes. Although only total hospital admissions are shown here, the fitted model distinguishes variant-specific hospitalization contributions through the latent infection process; corresponding variant-resolved results, together with replication analyses under alternative case ascertainment scenarios, are reported in Supplementary Material.

\subsubsection{Parameter estimates}

~\autoref{scenario_analysis} reports parameter estimates for the viral load model under the three case ascertainment scenarios. Estimates are shown for the two variant groups ($k=1$ pre-Omicron and $k=2$ Omicron), with results stratified by period indicators used as population-level covariates; overlap in listed variants (e.g., BA.1) reflects allowed co-circulation between the two groups. The reported mean and variance columns are derived quantities computed from the estimated distributional parameters and represent the average per infection daily contribution to the wastewater viral signal.

Across scenarios, variant-specific contrasts are consistent. In the pre-Omicron group, the Wild-type variant is associated with a larger inferred mean wastewater signal per infection relative to the Delta/BA.1 baseline, while Alpha is close to the baseline. In the Omicron group, BA.2 and BA.3 have substantially lower regression coefficients than BA.1, implying larger inferred mean signals on the original scale. These findings are directionally consistent with previous wastewater surveillance studies documenting higher SARS-CoV-2 shedding rates during Delta-dominant periods and lower rates during Omicron periods (e.g., \cite{prasek2023variant}), despite differences in estimands and population- versus individual-level interpretations.

Case-ascertainment assumptions mainly affect the overall scale of the estimated mean and variance parameters rather than the relative ordering of variants. As under-reporting assumptions become more restrictive (Scenario~1 to Scenario~3), the estimated per-infection wastewater signal decreases, while variant contrasts remain stable.

~\autoref{scenario_analysis_hosp} reports estimates of the hospitalization hazard parameters under the three case-ascertainment scenarios, together with the corresponding hazard rates for time to first hospitalization. The reported hazard rates are derived quantities computed from the estimated model parameters.

Across all scenarios, baseline hospitalization hazards are substantially lower for the Omicron variant group than for the pre-Omicron group. Within the Omicron group, BA.2 and BA.3 are consistently associated with higher hospitalization hazards than BA.1, with BA.3 exhibiting the highest hazard. In the pre-Omicron group, Alpha shows a higher hospitalization hazard relative to the Delta/BA.1 baseline, while differences for the Wild-type variant are smaller.

\section{Discussion}
\label{W3:dis}
This work develops a modeling framework for jointly analyzing wastewater viral signals and hospitalization data through a latent infection process, motivated by structural limitations of clinical surveillance and the growing role of wastewater-based infectious disease monitoring. Although the model is formulated at the individual level, inference is conducted using only aggregated population-level data, reflecting the structure of most real world surveillance systems. The primary contribution lies in demonstrating how such a framework can be operationalized to extract interpretable epidemiological signals in settings where infections are imperfectly observed and individual-level data are unavailable. In particular, the framework highlights how environmental surveillance can support inference on community-level disease burden when clinical case counts are systematically under-reported.

A central challenge addressed in this work is the absence of individual-level infection trajectories. By treating infection as an unobserved driving process and modeling wastewater signals and hospitalizations as conditionally independent given infection, the framework allows these complementary data sources to jointly inform transmission dynamics. This structure avoids imposing a direct mechanistic link between environmental surveillance and clinical metrics, which is unlikely to be stable over time, and instead attributes their association to evolving infection prevalence and variant composition.

A second major challenge is systematic under-reporting in aggregated clinical case counts data. Rather than treating reported clinical case counts as unbiased observations of infection prevalence, we explicitly account for incomplete ascertainment through a scenario based analysis informed by external information on testing accessibility. This formulation reflects a realistic surveillance setting in which complete clinical ascertainment is unlikely, particularly during the emergence of a novel pathogen, rather than a temporary limitation to be resolved by improved testing. By distinguishing periods of broad versus limited testing, the model incorporates reported clinical case counts as incomplete observations on the latent infection process, rather than as exact measurements. Both simulation results and the Ottawa case study demonstrate that this approach yields stable inference across a range of under-reporting assumptions, while preserving key temporal and variant-specific patterns.

Several limitations merit consideration. Because inference relies on aggregated data, the model cannot capture individual-level heterogeneity related to age, comorbidities, vaccination status, or prior immunity. Moreover, when infection prevalence is entirely unobserved, absolute levels of inferred infections are not identifiable without external information. While the proposed approach recovers relative temporal patterns and variant contrasts, incorporating additional data sources, such as seroprevalence studies or calibrated infection estimates, would be necessary to identify absolute infection burdens. A related limitation concerns uncertainty in case ascertainment. Further numerical studies could examine the efficiency and robustness of the proposed approach when assumptions on under-reporting are misspecified, which we leave for future work. An additional practical consideration concerns model diagnostics in applied surveillance settings. While formal diagnostic tools are important for assessing model adequacy, models intended for routine monitoring must operate under constraints on data availability and analytical complexity. Developing diagnostic approaches that balance statistical rigor with simplicity is therefore an important direction for future work when deploying latent infection models in public health monitoring systems.

From an applied perspective, the proposed framework is particularly well suited to surveillance settings characterized by incomplete or evolving clinical case ascertainment. Examples include early outbreak phases in which testing capacity is limited, as well as later-stage monitoring when routine case reporting is reduced or discontinued. The explicit modeling of a latent infection process further allows the framework to accommodate periods of co-circulation of multiple variants or strains. Although illustrated using COVID-19 data, the framework is broadly applicable to other infectious diseases for which wastewater surveillance provides informative population-level signals and clinical outcomes are observed only in aggregated form, with appropriate adaptations to reflect pathogen-specific transmission dynamics, shedding behavior, and available surveillance data.

Future work could extend this framework in several directions.  With richer data—whether from wastewater, serological studies, or other population-level sources—the infection and hospitalization processes could incorporate more flexible transition structures, delays, or covariate effects to better reflect clinical and behavioral heterogeneity. From a surveillance perspective, integrating real time wastewater monitoring with short term nowcasting or forecasting may further enhance its utility for public health decision making, particularly in periods of limited or rapidly changing clinical testing capacity. More broadly, developing inference strategies for latent infection processes under severe aggregation and under-reporting remains an important area for methodological research.

\section*{Acknowledgments}
This research is partially supported by the grants of Natural Sciences and Engineering Research Council of Canada (NSERC) to Hu and Swartz, and the CRT (Collaborative Research Team) in Sports Analytics of the Canadian Statistical Sciences Institute (CANSSI) led by Swartz.
The authors thank Daniel Stenz, former Technical Director of Shandong Luneng Taishan FC for providing the data discussed in this paper.

\section*{Declarations}
The authors declare no conflict of interest.

\clearpage

\appendix

\section{Figures}

\clearpage

\section{Tables}

\begin{table}[!ht]
\centering
\footnotesize
\noindent\textit{(a) $(r_1, r_2) = (0.8, 0.8)$}\par
\begin{tabular}{l | c | c c | c c c c}
\toprule
& \textbf{True} 
& \multicolumn{2}{c|}{\textbf{Ignore under-reporting}} 
& \multicolumn{4}{c}{\textbf{Proposed approach}} \\
\cmidrule(lr){2-2} \cmidrule(lr){3-4} \cmidrule(lr){5-8}
Param 
&  
& $\overline{\text{EST}}_{\text{naive}}$ 
& $SD(\text{EST}_{\text{naive}})$ 
& $\overline{\text{EST}}$ 
& $SD(\text{EST})$ 
& $\overline{\hat{SE}_1}$ 
& $\overline{\hat{SE}_2}$ \\
\midrule
$\lambda_1$ ($10^{-3}$) & 2.00 & 2.23 & 0.38 & 1.95 & 0.14 & 0.07 & 0.14 \\
$\lambda_2$ ($10^{-3}$) & 5.00 & 5.44 & 1.25 & 4.55 & 0.25 & 0.13 & 0.29 \\
$\alpha_1$ ($10^{-3}$)  & 1.00 & 1.68 & 1.59 & 0.96 & 0.41 & 0.09 & 0.34 \\
$\beta_1$ ($10^{4}$)    & 1.00 & 1.69 & 1.66 & 0.98 & 0.49 & 0.10 & 0.41 \\
$\alpha_2$ ($10^{-3}$)  & 5.00 & 5.24 & 4.88 & 4.72 & 1.45 & 0.53 & 1.20 \\
$\beta_2$ ($10^{4}$)    & 2.00 & 2.22 & 1.96 & 1.92 & 0.63 & 0.22 & 0.59 \\
\bottomrule
\end{tabular}

\vspace{1em}

\noindent\textit{(b) $(r_1, r_2) = (0.2, 0.8)$}\par
\begin{tabular}{l | c | c c | c c c c}
\toprule
& \textbf{True} 
& \multicolumn{2}{c|}{\textbf{Ignore under-reporting}} 
& \multicolumn{4}{c}{\textbf{Proposed approach}} \\
\cmidrule(lr){2-2} \cmidrule(lr){3-4} \cmidrule(lr){5-8}
Param 
&  
& $\overline{\text{EST}}_{\text{naive}}$ 
& $SD(\text{EST}_{\text{naive}})$ 
& $\overline{\text{EST}}$ 
& $SD(\text{EST})$ 
& $\overline{\hat{SE}_1}$ 
& $\overline{\hat{SE}_2}$ \\
\midrule
$\lambda_1$ ($10^{-3}$) & 2.00 & 2.66 & 0.24 & 1.96 & 0.14 & 0.07 & 0.15 \\
$\lambda_2$ ($10^{-3}$) & 5.00 & 5.67 & 0.51 & 4.51 & 0.26 & 0.13 & 0.32 \\
$\alpha_1$ ($10^{-3}$)  & 1.00 & 1.06 & 0.35 & 0.96 & 0.45 & 0.11 & 0.42 \\
$\beta_1$ ($10^{4}$)    & 1.00 & 0.92 & 0.33 & 1.04 & 0.52 & 0.13 & 0.50 \\
$\alpha_2$ ($10^{-3}$)  & 5.00 & 4.91 & 1.89 & 4.86 & 1.86 & 0.55 & 1.32 \\
$\beta_2$ ($10^{4}$)    & 2.00 & 1.69 & 0.64 & 1.95 & 0.77 & 0.24 & 0.66 \\
\bottomrule
\end{tabular}

\vspace{1em}

\noindent\textit{(c) $(r_1, r_2) = (0.8, 0.2)$}\par
\begin{tabular}{l | c | c c | c c c c}
\toprule
& \textbf{True} 
& \multicolumn{2}{c|}{\textbf{Ignore under-reporting}} 
& \multicolumn{4}{c}{\textbf{Proposed approach}} \\
\cmidrule(lr){2-2} \cmidrule(lr){3-4} \cmidrule(lr){5-8}
Param 
&  
& $\overline{\text{EST}}_{\text{naive}}$ 
& $SD(\text{EST}_{\text{naive}})$ 
& $\overline{\text{EST}}$ 
& $SD(\text{EST})$ 
& $\overline{\hat{SE}_1}$ 
& $\overline{\hat{SE}_2}$ \\
\midrule
$\lambda_1$ ($10^{-3}$) & 2.00 & 2.55 & 0.25 & 2.08 & 0.20 & 0.07 & 0.17 \\
$\lambda_2$ ($10^{-3}$) & 5.00 & 5.52 & 0.31 & 4.78 & 0.38 & 0.13 & 0.31 \\
$\alpha_1$ ($10^{-3}$)  & 1.00 & 1.23 & 0.34 & 1.04 & 0.36 & 0.10 & 0.41 \\
$\beta_1$ ($10^{4}$)    & 1.00 & 1.02 & 0.31 & 1.03 & 0.38 & 0.12 & 0.44 \\
$\alpha_2$ ($10^{-3}$)  & 5.00 & 4.74 & 1.04 & 4.80 & 1.52 & 0.51 & 1.46 \\
$\beta_2$ ($10^{4}$)    & 2.00 & 1.62 & 0.36 & 1.89 & 0.63 & 0.22 & 0.57 \\
\bottomrule
\end{tabular}

\vspace{1em}

\noindent\textit{(d) $(r_1, r_2) = (0.2, 0.2)$}\par
\begin{tabular}{l | c | c c | c c c c}
\toprule
& \textbf{True} 
& \multicolumn{2}{c|}{\textbf{Ignore under-reporting}} 
& \multicolumn{4}{c}{\textbf{Proposed approach}} \\
\cmidrule(lr){2-2} \cmidrule(lr){3-4} \cmidrule(lr){5-8}
Param 
&  
& $\overline{\text{EST}}_{\text{naive}}$ 
& $SD(\text{EST}_{\text{naive}})$ 
& $\overline{\text{EST}}$ 
& $SD(\text{EST})$ 
& $\overline{\hat{SE}_1}$ 
& $\overline{\hat{SE}_2}$ \\
\midrule
$\lambda_1$ ($10^{-3}$) & 2.00 & 5.99  & 0.51 & 2.20 & 0.22 & 0.07 & 0.22 \\
$\lambda_2$ ($10^{-3}$) & 5.00 & 13.06 & 0.92 & 4.93 & 0.41 & 0.14 & 0.33 \\
$\alpha_1$ ($10^{-3}$)  & 1.00 & 2.78  & 1.30 & 1.04 & 0.34 & 0.12 & 0.32 \\
$\beta_1$ ($10^{4}$)    & 1.00 & 1.14  & 0.60 & 0.96 & 0.37 & 0.13 & 0.32 \\
$\alpha_2$ ($10^{-3}$)  & 5.00 & 2.09  & 1.60 & 5.07 & 1.82 & 0.58 & 1.49 \\
$\beta_2$ ($10^{4}$)    & 2.00 & 0.39  & 0.31 & 2.09 & 0.69 & 0.23 & 0.52 \\
\bottomrule
\end{tabular}
\caption{Simulation results under four $(r_1, r_2)$ scenarios in Section~\ref{w3:simulation:result}.}
\label{tabw3:simulation}
\end{table}

\begin{table}
    \centering
    \footnotesize
    \begin{tabular}{lrrr}
        \toprule
        \textbf{Periods} & \textbf{$\pmb{\beta}$, Est (SE)} & \textbf{Mean, Est (SE)} & \textbf{Var, Est (SE)} \\
        \midrule
        \multicolumn{4}{l}{\textbf{Scenario 1: consider observed $S_t$ are true number of infections}} \\
        \midrule
        \multicolumn{4}{l}{\textbf{\textit{Pre-Omicron periods (shape: 0.004 (0.0002))}}} \\
        Intercept $\beta_0$: & $9.57 (0.094)$ &  &  \\
        Delta \& BA.1 (baseline) & $\cdot$ & $28.1 (2.21) \times 10^{-8}$ & $19.5 (3.22) \times 10^{-12}$ \\
        Wild & $\pmb{-0.51 (0.086)}$ & $46.7 (1.65) \times 10^{-8}$ & $54.0 (4.68) \times 10^{-12}$ \\
        Alpha & $-0.01 (0.088)$ & $27.8 (1.07) \times 10^{-8}$ & $19.1 (1.77) \times 10^{-12}$ \\
        \midrule
        \multicolumn{4}{l}{\textbf{\textit{Omicron periods (shape: 0.001 (0.0001))}}} \\
        Intercept $\beta_0$: & $9.79 (0.079)$ &  &  \\
        BA.1 (baseline) & $\cdot$ & $59.7 (2.69) \times 10^{-9}$ & $33.3 (3.71) \times 10^{-13}$ \\
        BA.2 & $\pmb{-2.27 (0.101)}$ & $57.5 (5.17) \times 10^{-8}$ & $31.0 (5.91) \times 10^{-11}$ \\
        BA.3 & $\pmb{-2.35 (0.088)}$ & $62.7 (4.74) \times 10^{-8}$ & $36.8 (6.05) \times 10^{-11}$ \\
        \midrule
        \multicolumn{4}{l}{\textbf{Scenario 2: consider $L(t)$}} \\
        \midrule
        \multicolumn{4}{l}{\textbf{\textit{Pre-Omicron periods (shape: 0.004 (0.0002))}}} \\
        Intercept $\beta_0$: & $9.63 (0.092)$ &  &  \\
        Delta \& BA.1 (baseline) & $\cdot$ & $26.3 (2.02) \times 10^{-8}$ & $17.3 (2.80) \times 10^{-12}$ \\
        Wild & $\pmb{-0.51 (0.084)}$ & $44.0 (1.51) \times 10^{-8}$ & $48.3 (4.11) \times 10^{-12}$ \\
        Alpha & $-0.05 (0.084)$ & $27.8 (1.08) \times 10^{-8}$ & $19.2 (1.78) \times 10^{-12}$ \\
        \midrule
        \multicolumn{4}{l}{\textbf{\textit{Omicron periods (shape: 0.001 (0.0001))}}} \\
        Intercept $\beta_0$: & $9.99 (0.078)$ &  &  \\
        BA.1 (baseline) & $\cdot$ & $45.6 (1.86) \times 10^{-9}$ & $20.8 (2.20) \times 10^{-13}$ \\
        BA.2 & $\pmb{-2.16 (0.088)}$ & $39.6 (3.06) \times 10^{-8}$ & $15.7 (2.64) \times 10^{-11}$ \\
        BA.3 & $\pmb{-2.17 (0.075)}$ & $40.0 (2.51) \times 10^{-8}$ & $16.1 (2.28) \times 10^{-11}$ \\
        \midrule
        \multicolumn{4}{l}{\textbf{Scenario 3: consider $S_t$ are always lower than true infections}} \\
        \midrule
        \multicolumn{4}{l}{\textbf{\textit{Pre-Omicron periods (shape: 0.003 (0.0001))}}} \\
        Intercept $\beta_0$: & $9.97 (0.083)$ &  &  \\
        Delta \& BA.1 (baseline) & $\cdot$ & $12.4 (0.80) \times 10^{-8}$ & $5.79 (0.81) \times 10^{-12}$ \\
        Wild & $\pmb{-1.06 (0.075)}$ & $35.8 (1.36) \times 10^{-8}$ & $48.3 (4.47) \times 10^{-12}$ \\
        Alpha & $\pmb{-0.55 (0.077)}$ & $21.4 (0.90) \times 10^{-8}$ & $17.2 (1.71) \times 10^{-12}$ \\
        \midrule
        \multicolumn{4}{l}{\textbf{\textit{Omicron periods (shape: 0.001 (0.0001))}}} \\
        Intercept $\beta_0$: & $10.17 (0.077)$ &  &  \\
        BA.1 (baseline) & $\cdot$ & $46.5 (1.73) \times 10^{-9}$ & $17.7 (1.77) \times 10^{-13}$ \\
        BA.2 & $\pmb{-2.14 (0.079)}$ & $39.5 (2.76) \times 10^{-8}$ & $12.8 (1.98) \times 10^{-11}$ \\
        BA.3 & $\pmb{-2.17 (0.068)}$ & $40.8 (2.33) \times 10^{-8}$ & $13.6 (1.80) \times 10^{-11}$ \\
        \bottomrule
    \end{tabular}
    \caption{Comparison of parameter estimates under three case-ascertainment scenarios. 
The first column reports estimates of the regression coefficients $\pmb{\beta}$ with standard errors in parentheses. 
The second column reports the estimated mean of the variant-specific gamma distribution, obtained by plugging in the estimated shape and rate parameters, with standard errors computed via the multivariate delta method. 
The third column reports the corresponding estimated variance of the gamma distribution, with standard errors computed analogously. Scenario~1 assumes reported active cases $S_t$ equal the true number of infections; Scenario~2 accounts for time-varying testing accessibility through $L(t)$; and Scenario~3 assumes persistent under-reporting of cases throughout the study period.}
    \label{scenario_analysis}
\end{table}

\begin{table}
    \centering
    \footnotesize
    \begin{tabular}{lrc}
        \toprule
        \textbf{Periods} & \textbf{$\pmb{\lambda}$, Est (SE)} & \textbf{Hazard, Est (SE)} \\
        \midrule
        \multicolumn{3}{l}{\textbf{Scenario 1}} \\
                \midrule
        \multicolumn{3}{l}{\textbf{\textit{Pre-Omicron periods}}} \\
        Intercept $\lambda_0$: & \multicolumn{1}{r}{$-5.77\ (0.090)$} &  \\
        Delta \& BA.1 (baseline) & \multicolumn{1}{c}{$\cdot$} & $0.0031\ (0.0003)$ \\
        Wild                  & $-0.063\ (0.098)$ & $0.0029\ (0.0001)$ \\
        Alpha                    & $\pmb{0.302\ (0.097)}$ & $0.0042\ (0.0002)$ \\
        \midrule
        \multicolumn{3}{l}{\textbf{\textit{Omicron periods}}} \\
        Intercept $\lambda_0$: & \multicolumn{1}{r}{$-6.53\ (0.038)$} &  \\
        BA.1 (baseline)          & \multicolumn{1}{c}{$\cdot$} & $0.0015\ (0.0001)$ \\
        BA.2                     & $\pmb{0.437\ (0.073)}$ & $0.0023\ (0.0001)$ \\
        BA.3                     & $\pmb{1.146\ (0.053)}$ & $0.0046\ (0.0002)$ \\
        \midrule
        \multicolumn{3}{l}{\textbf{Scenario 2}} \\
        \midrule
        \multicolumn{3}{l}{\textbf{\textit{Pre-Omicron periods}}} \\
        Intercept $\lambda_0$: & \multicolumn{1}{r}{$-5.74\ (0.084)$} &  \\
        Delta \& BA.1 (baseline) & \multicolumn{1}{c}{$\cdot$} & $0.0032\ (0.0003)$ \\
        Wild                  & $-0.14\ (0.095)$ & $0.0028\ (0.0001)$ \\
        Alpha                    & $\pmb{0.283\ (0.094)}$ & $0.0042\ (0.0002)$ \\
        \midrule
        \multicolumn{3}{l}{\textbf{\textit{Omicron periods}}} \\
        Intercept $\lambda_0$: & \multicolumn{1}{r}{$-6.81\ (0.039)$} &  \\
        BA.1 (baseline)          & \multicolumn{1}{c}{$\cdot$} & $0.0011\ (0.0001)$ \\
        BA.2                     & $\pmb{0.356\ (0.073)}$ & $0.0016\ (0.0001)$ \\
        BA.3                     & $\pmb{0.989\ (0.053)}$ & $0.0029\ (0.0001)$ \\
        \midrule
        \multicolumn{3}{l}{\textbf{Scenario 3}} \\
        \midrule
        \multicolumn{3}{l}{\textbf{\textit{Pre-Omicron periods}}} \\
        Intercept $\lambda_0$: & \multicolumn{1}{r}{$-6.27\ (0.076)$} &  \\
        Delta \& BA.1 (baseline) & \multicolumn{1}{c}{$\cdot$} & $0.0019\ (0.0001)$ \\
        Wild                  & $0.171\ (0.086)$ & $0.0022\ (0.0001)$ \\
        Alpha                    & $\pmb{0.541\ (0.085)}$ & $0.0033\ (0.0001)$ \\
        \midrule
        \multicolumn{3}{l}{\textbf{\textit{Omicron periods}}} \\
        Intercept $\lambda_0$: & \multicolumn{1}{r}{$-6.85\ (0.040)$} &  \\
        BA.1 (baseline)          & \multicolumn{1}{c}{$\cdot$} & $0.0011\ (0.0001)$ \\
        BA.2                     & $\pmb{0.384\ (0.074)}$ & $0.0016\ (0.0001)$ \\
        BA.3                     & $\pmb{1.038\ (0.054)}$ & $0.0030\ (0.0001)$ \\
        \bottomrule
    \end{tabular}
    \caption{Comparison of hospitalization intensity parameter estimates under three case-ascertainment scenarios. The first column reports estimates of the regression coefficients $\pmb{\lambda}$ with standard errors in parentheses. The second column reports the corresponding estimated hazards of time to hospital, obtained by evaluating the fitted intensity function at the baseline covariate values, with standard errors derived via the delta method.}
    \label{scenario_analysis_hosp}
\end{table}

\clearpage

\bibliographystyle{agsm}
\bibliography{Bibliography-MM-MC}

@article{peng2023exploration,
  title={An exploration of the relationship between wastewater viral signals and COVID-19 hospitalizations in Ottawa, Canada},
  author={Peng, K Ken and Renouf, Elizabeth M and Dean, Charmaine B and Hu, X Joan and Delatolla, Robert and Manuel, Douglas G},
  journal={Infectious Disease Modelling},
  volume={8},
  number={3},
  pages={617--631},
  year={2023},
  publisher={Elsevier}
}

@article{prasek2023variant,
  title={Variant-specific SARS-CoV-2 shedding rates in wastewater},
  author={Prasek, Sarah M and Pepper, Ian L and Innes, Gabriel K and Slinski, Stephanie and Betancourt, Walter Q and Foster, Aidan R and Yaglom, Hayley D and Porter, W Tanner and Engelthaler, David M and Schmitz, Bradley W},
  journal={Science of The Total Environment},
  volume={857},
  pages={159165},
  year={2023},
  publisher={Elsevier}
}

@article{gilani2017anthropometric,
  title={Anthropometric indices as predictors of coronary heart disease risk: Joint modeling of longitudinal measurements and time to event},
  author={Gilani, Neda and Kazemnejad, Anoshirvan and Zayeri, Farid and Hadaegh, Farzad and Azizi, Fereidoun and Khalili, Davood},
  journal={Iranian journal of public health},
  volume={46},
  number={11},
  pages={1546},
  year={2017}
}

@article{zelelew2023joint,
  title={Joint Modeling of Blood Pressure Measurements and Survival Time to Cardiovascular Disease Complication among Hypertension Patients Follow-up at DebreTabor Hospital, Ethiopia},
  author={Zelelew, Abebe Nega and Workie, Demeke Lakew},
  journal={Vascular Health and Risk Management},
  pages={621--635},
  year={2023},
  publisher={Taylor \& Francis}
}

@article{luvanda2023joint,
  title={A joint survival model for estimating the association between viral load outcome and survival time to death among HIV/AIDS patients attending health care and treatment centers in Tanzania},
  author={Luvanda, Habiel Benjamin and Mukyanuzi, Elevatus Nkebukwa and Akarro, Rocky RJ},
  journal={BMC Public Health},
  volume={23},
  number={1},
  pages={2091},
  year={2023},
  publisher={Springer}
}

@article{temesgen2018joint,
  title={Joint modeling of longitudinal CD4 count and time-to-death of HIV/TB co-infected patients: A case of Jimma University Specialized Hospital},
  author={Temesgen, Aboma and Gurmesa, Abdisa and Getchew, Yehenew},
  journal={Annals of Data Science},
  volume={5},
  number={4},
  pages={659--678},
  year={2018},
  publisher={Springer}
}

@article{wu2012analysis,
  title={Analysis of longitudinal and survival data: joint modeling, inference methods, and issues},
  author={Wu, Lang and Liu, Wei and Yi, Grace Y and Huang, Yangxin},
  journal={Journal of Probability and Statistics},
  volume={2012},
  number={1},
  pages={640153},
  year={2012},
  publisher={Wiley Online Library}
}

@article{tsiatis2004joint,
  title={Joint modeling of longitudinal and time-to-event data: an overview},
  author={Tsiatis, Anastasios A and Davidian, Marie},
  journal={Statistica Sinica},
  pages={809--834},
  year={2004},
  publisher={JSTOR}
}

@article{wu2010joint,
  title={Joint inference on HIV viral dynamics and immune suppression in presence of measurement errors},
  author={Wu, L and Liu, W and Hu, XJ},
  journal={Biometrics},
  volume={66},
  number={2},
  pages={327--335},
  year={2010},
  publisher={Oxford University Press}
}

@article{albani2021covid,
  title={COVID-19 underreporting and its impact on vaccination strategies},
  author={Albani, Vinicius and Loria, Jennifer and Massad, Eduardo and Zubelli, Jorge},
  journal={BMC infectious diseases},
  volume={21},
  pages={1--13},
  year={2021},
  publisher={Springer}
}

@article{lau2021evaluating,
  title={Evaluating the massive underreporting and undertesting of COVID-19 cases in multiple global epicenters},
  author={Lau, Hien and Khosrawipour, Tanja and Kocbach, Piotr and Ichii, Hirohito and Bania, Jacek and Khosrawipour, Veria},
  journal={Pulmonology},
  volume={27},
  number={2},
  pages={110--115},
  year={2021},
  publisher={Elsevier}
}

@article{miyazawa2024wastewater,
  title={Wastewater-based reproduction numbers and projections of COVID-19 cases in three areas in Japan, November 2021 to December 2022},
  author={Miyazawa, Shogo and Wong, Ting Sam and Ito, Genta and Iwamoto, Ryo and Watanabe, Kozo and van Boven, Michiel and Wallinga, Jacco and Miura, Fuminari},
  journal={Eurosurveillance},
  volume={29},
  number={8},
  pages={2300277},
  year={2024}
}

@article{meadows2025epidemiological,
  title={Epidemiological model can forecast COVID-19 outbreaks from wastewater-based surveillance in rural communities},
  author={Meadows, Tyler and Coats, Erik R and Narum, Solana and Top, Eva M and Ridenhour, Benjamin J and Stalder, Thibault},
  journal={Water Research},
  volume={268},
  pages={122671},
  year={2025},
  publisher={Elsevier}
}

@article{simone2024time,
  title={From time series to visibility algorithms: A novel approach to study the spread of SARS-CoV-2 in wastewater},
  author={Simone, A and Cesaro, A and Esposito, G},
  journal={Journal of Water Process Engineering},
  volume={66},
  pages={106107},
  year={2024},
  publisher={Elsevier}
}

@article{lai2025temporal,
  title={Temporal cross-validation in forecasting: A case study of COVID-19 incidence using wastewater data},
  author={Lai, Mallory and Wulff, Shaun S and Cao, Yongtao and Robinson, Timothy J and Rajapaksha, Rasika},
  journal={Quality and Reliability Engineering International},
  volume={41},
  number={2},
  pages={672--688},
  year={2025},
  publisher={Wiley Online Library}
}

@article{jeng2025forecasting,
  title={Forecasting COVID-19 Cases, Hospital Admissions, and Deaths Based on Wastewater SARS-CoV-2 Surveillance Using Gaussian Copula Time Series Marginal Regression Model},
  author={Jeng, Hueiwang Anna and Diawara, Norou and Welch, Nancy and Jackson, Cynthia and Singh, Rekha and Curtis, Kyle and Gonzalez, Raul and Jurgens, David and Adikari, Sasanka},
  journal={COVID},
  volume={5},
  number={2},
  pages={25},
  year={2025},
  publisher={MDPI}
}

@article{starnini2021impact,
  title={Impact of data accuracy on the evaluation of COVID-19 mitigation policies},
  author={Starnini, Michele and Aleta, Alberto and Tizzoni, Michele and Moreno, Yamir},
  journal={Data \& Policy},
  volume={3},
  pages={e28},
  year={2021},
  publisher={Cambridge University Press}
}

@article{pappu2025tracking,
  title={Tracking COVID-19 trends in communities with low population by wastewater-based surveillance},
  author={Pappu, Aiswarya Rani and Green, Ashley and Oakes, Melanie and Jiang, Sunny},
  journal={Science of The Total Environment},
  volume={970},
  pages={179007},
  year={2025},
  publisher={Elsevier}
}

@article{mohring2024estimating,
  title={Estimating the COVID-19 prevalence from wastewater},
  author={Mohring, Jan and Leith{\"a}user, Neele and Wlaz{\l}o, Jaros{\l}aw and Schulte, Marvin and Pilz, Maximilian and M{\"u}nch, Johanna and K{\"u}fer, Karl-Heinz},
  journal={Scientific Reports},
  volume={14},
  number={1},
  pages={14384},
  year={2024},
  publisher={Nature Publishing Group UK London}
}

@article{holcomb2024estimating,
  title={Estimating rates of change to interpret quantitative wastewater surveillance of disease trends},
  author={Holcomb, David A and Christensen, Ariel and Hoffman, Kelly and Lee, Allison and Blackwood, A Denene and Clerkin, Thomas and Gallard-G{\'o}ngora, Javier and Harris, Angela and Kotlarz, Nadine and Mitasova, Helena and others},
  journal={Science of The Total Environment},
  volume={951},
  pages={175687},
  year={2024},
  publisher={Elsevier}
}

@article{parkins2024wastewater,
  title={Wastewater-based surveillance as a tool for public health action: SARS-CoV-2 and beyond},
  author={Parkins, Michael D and Lee, Bonita E and Acosta, Nicole and Bautista, Maria and Hubert, Casey RJ and Hrudey, Steve E and Frankowski, Kevin and Pang, Xiao-Li},
  journal={Clinical microbiology reviews},
  volume={37},
  number={1},
  pages={e00103--22},
  year={2024},
  publisher={American Society for Microbiology 1752 N St., NW, Washington, DC}
}

@article{nguyen2023estimation,
  title={Estimation of parameter distributions for reaction-diffusion equations with competition using aggregate spatiotemporal data},
  author={Nguyen, Kyle and Rutter, Erica M and Flores, Kevin B},
  journal={Bulletin of Mathematical Biology},
  volume={85},
  number={7},
  pages={62},
  year={2023},
  publisher={Springer}
}

@article{banks2020parameter,
  title={Parameter estimation using aggregate data},
  author={Banks, HT and Meade, Annabel E and Schacht, Celia and Catenacci, Jared and Thompson, W Clayton and Abate-Daga, Daniel and Enderling, Heiko},
  journal={Applied Mathematics Letters},
  volume={100},
  pages={105999},
  year={2020},
  publisher={Elsevier}
}

@article{PENG2025100840,
title = {Learning associations of COVID-19 hospitalizations with wastewater viral signals by Markov modulated models},
journal = {Epidemics},
volume = {52},
pages = {100840},
year = {2025},
issn = {1755-4365},
author = {K. Ken Peng and Charmaine B. Dean and X. Joan Hu and Robert Delatolla}
}

@article{orcutt1968data,
  title={Data aggregation and information loss},
  author={Orcutt, Guy H and Watts, Harold W and Edwards, John B},
  journal={The American Economic Review},
  volume={58},
  number={4},
  pages={773--787},
  year={1968},
  publisher={JSTOR}
}

@article{kelton1987comparison,
  title={Comparison of hypothesis testing techniques for markov processes estimated from micro versus macro data},
  author={Kelton, Christina ML and Kelton, W David},
  journal={Annals of Operations Research},
  volume={8},
  number={1},
  pages={175--194},
  year={1987},
  publisher={Springer}
}

@article{lawless1984information,
  title={The information in aggregate data from Markov chains},
  author={Lawless, JF and McLeish, DL},
  journal={Biometrika},
  volume={71},
  number={3},
  pages={419--430},
  year={1984},
  publisher={Oxford University Press}
}

@article{davis2002estimating,
  title={Estimating and interpolating a Markov chain from aggregate data},
  author={Davis, BA and Heathcote, Christopher R and O'neill, TJ},
  journal={Biometrika},
  volume={89},
  number={1},
  pages={95--110},
  year={2002},
  publisher={Oxford University Press}
}

@article{chen2017estimation,
  title={Estimation of field reliability based on aggregate lifetime data},
  author={Chen, Piao and Ye, Zhi-Sheng},
  journal={Technometrics},
  volume={59},
  number={1},
  pages={115--125},
  year={2017},
  publisher={Taylor \& Francis}
}

@article{sherratt2021exploring,
  title={Exploring surveillance data biases when estimating the reproduction number: with insights into subpopulation transmission of COVID-19 in England},
  author={Sherratt, Katharine and Abbott, Sam and Meakin, Sophie R and Hellewell, Joel and Munday, James D and Bosse, Nikos and CMMID Covid-19 working group and Jit, Mark and Funk, Sebastian},
  journal={Philosophical Transactions of the Royal Society B},
  volume={376},
  number={1829},
  pages={20200283},
  year={2021},
  publisher={The Royal Society}
}

@article{peccia2020measurement,
  title={{Measurement of SARS-CoV-2 RNA in wastewater tracks community infection dynamics}},
  author={Peccia, Jordan and Zulli, Alessandro and Brackney, Doug E and Grubaugh, Nathan D and Kaplan, Edward H and Casanovas-Massana, Arnau and Ko, Albert I and Malik, Amyn A and Wang, Dennis and Wang, Mike and others},
  journal={Nature Biotechnology},
  volume={38},
  number={10},
  pages={1164--1167},
  year={2020},
  publisher={Nature Publishing Group}
}

@article{galani2022sars,
  title={{SARS-CoV-2 wastewater surveillance data can predict hospitalizations and ICU admissions}},
  author={Galani, Aikaterini and Aalizadeh, Reza and Kostakis, Marios and Markou, Athina and Alygizakis, Nikiforos and Lytras, Theodore and Adamopoulos, Panagiotis G and Peccia, Jordan and Thompson, David C and Kontou, Aikaterini and others},
  journal={Science of The Total Environment},
  volume={804},
  pages={150151},
  year={2022},
  publisher={Elsevier}
}

@article{he2020temporal,
  title={Temporal dynamics in viral shedding and transmissibility of {COVID}-19},
  author={He, Xi and Lau, Eric HY and Wu, Peng and Deng, Xilong and Wang, Jian and Hao, Xinxin and Lau, Yiu Chung and Wong, Jessica Y and Guan, Yujuan and Tan, Xinghua and others},
  journal={Nature Medicine},
  volume={26},
  number={5},
  pages={672--675},
  year={2020},
  publisher={Nature Publishing Group US New York}
}

@article{graber2021near,
  title={Near real-time determination of B. 1.1. 7 in proportion to total SARS-CoV-2 viral load in wastewater using an allele-specific primer extension PCR strategy},
  author={Graber, Tyson E and Mercier, {\'E}lisabeth and Bhatnagar, Kamya and Fuzzen, Meghan and D'Aoust, Patrick M and Hoang, Huy-Dung and Tian, Xin and Towhid, Syeda Tasneem and Plaza-Diaz, Julio and Eid, Walaa and others},
  journal={Water research},
  volume={205},
  pages={117681},
  year={2021},
  publisher={Elsevier}
}

@article{kaplan2021aligning,
  title={{Aligning SARS-CoV-2 indicators via an epidemic model: application to hospital admissions and RNA detection in sewage sludge}},
  author={Kaplan, Edward H and Wang, Dennis and Wang, Mike and Malik, Amyn A and Zulli, Alessandro and Peccia, Jordan},
  journal={Health Care Management Science},
  volume={24},
  number={2},
  pages={320--329},
  year={2021},
  publisher={Springer}
}

@article{zulli2022predicting,
  title={{Predicting daily COVID-19 case rates from SARS-CoV-2 RNA concentrations across a diversity of wastewater catchments}},
  author={Zulli, Alessandro and Pan, Annabelle and Bart, Stephen M and Crawford, Forrest W and Kaplan, Edward H and Cartter, Matthew and Ko, Albert I and Sanchez, Marcela and Brown, Cade and Cozens, Duncan and others},
  journal={FEMS microbes},
  volume={2},
  year={2022},
  publisher={Oxford Academic}
}

@article{schoen2022sars,
  title={{SARS-CoV-2 RNA wastewater settled solids surveillance frequency and impact on predicted COVID-19 incidence using a distributed lag model}},
  author={Schoen, Mary E and Wolfe, Marlene K and Li, Linlin and Duong, Dorothea and White, Bradley J and Hughes, Bridgette and Boehm, Alexandria B},
  journal={ACS ES\&T Water},
  year={2022},
  publisher={ACS Publications}
}

@article{xie2022rna,
  title={{RNA in Municipal Wastewater Reveals Magnitudes of COVID-19 Outbreaks across Four Waves Driven by SARS-CoV-2 Variants of Concern}},
  author={Xie, Yuwei and Challis, Jonathan K and Oloye, Femi F and Asadi, Mohsen and Cantin, Jenna and Brinkmann, Markus and McPhedran, Kerry N and Hogan, Natacha and Sadowski, Mike and Jones, Paul D and others},
  journal={ACS ES\&T Water},
  year={2022},
  publisher={ACS Publications}
}

@article{hegazy2022understanding,
  title={{Understanding the dynamic relation between wastewater SARS-CoV-2 signal and clinical metrics throughout the pandemic}},
  author={Hegazy, Nada and Cowan, Aaron and D'Aoust, Patrick M and Mercier, Elisabeth and Towhid, Syeda Tasneem and Jia, Jian-Jun and Wan, Shen and Zhang, Zhihao and Kabir, Md Pervez and Fang, Wanting and others},
  journal={medRxiv},
  year={2022},
  publisher={Cold Spring Harbor Laboratory Press}
}

@article{larsen2020tracking,
  title={{Tracking COVID-19 with wastewater}},
  author={Larsen, David A and Wigginton, Krista R},
  journal={Nature Biotechnology},
  volume={38},
  number={10},
  pages={1151--1153},
  year={2020},
  publisher={Nature Publishing Group}
}

@article{d2022wastewater,
  title={{Wastewater to clinical case (WC) ratio of COVID-19 identifies insufficient clinical testing, onset of new variants of concern and population immunity in urban communities}},
  author={D'Aoust, Patrick M and Tian, Xin and Towhid, Syeda Tasneem and Xiao, Amy and Mercier, Elisabeth and Hegazy, Nada and Jia, Jian-Jun and Wan, Shen and Kabir, Md Pervez and Fang, Wanting and others},
  journal={Science of The Total Environment},
  volume={853},
  pages={158547},
  year={2022},
  publisher={Elsevier}
}

\end{document}